\documentclass[twocolumn,showpacs,preprintnumbers,amsmath,amssymb]{revtex4}

\usepackage{graphicx}
\usepackage{dcolumn}
\usepackage{bm}
\usepackage{wrapfig} 

\begin{document}


\title{A Study of $e^+e^- \to H^0A^0$ Production  \\
and the Constraint on Dark Matter Density}

\author{Marco Battaglia}
 \email{MBattaglia@lbl.gov}
\affiliation{University of California at Berkeley, Department of Physics and\\ 
Lawrence Berkeley National Laboratory, Berkeley, CA - USA}%
\author{Benjamin Hooberman}%
 \email{benhooberman@berkeley.edu}
\affiliation{University of California at Berkeley, Department of Physics and\\ 
Lawrence Berkeley National Laboratory, Berkeley, CA - USA}%
\author{Nicole Kelley}%
 \email{kelley@berkeley.edu}
\affiliation{University of California at Berkeley, Department of Physics and\\
Lawrence Berkeley National Laboratory, Berkeley, CA - USA}%

\date{\today}

\begin{abstract}
This paper reports the results of a study of the $e^+e^- \to H^0A^0$ 
process at $\sqrt{s}$ = 1~TeV performed on fully simulated and reconstructed
events. The estimated accuracies on the heavy Higgs boson masses, widths and 
decay branching fractions are discussed in relation to the study of 
Supersymmetric Dark Matter.
\end{abstract}

\pacs{13.66.Fg, 14.80.Cp}
\maketitle

\section{Introduction}

The connections between cosmology and particle physics through dark datter
(DM) have recently received special attention for defining the physics 
program at the TeV frontier. We foresee that the combination of data from 
satellites, direct DM searches, hadron and lepton colliders will provide a
major breakthrough in our understanding of the nature of dark matter and
its interactions in the early Universe. These expectations are supported by 
the fact that there are several extensions of the Standard Model (SM), which 
include a new, stable, weakly-interacting massive particle, which may be 
responsible for the observed relic DM in the Universe. This particle should 
become accessible to particle colliders operating at the TeV energy frontier, 
as well as to the next generations of direct DM search experiments. 
The LHC collider will be first in providing data to address the 
question of whether one of these scenarios is indeed realised in nature. 
If this is the case, it will also gather some quantitative information to be 
related to the relic DM density measured from the cosmic microwave background 
(CMB) spectra~\cite{Battaglia:2004mp}. However, it is understood that the LHC 
data will not be exhaustive in this respect. First, it will not be possible to infer, 
in a model independent way, the relic density to an accuracy close to that already 
achieved by CMB observations. Furthermore, there exist classes of models of new 
physics which the LHC may not be able to disentangle and probe in sufficient details. 
It is only with the measurements becoming available at an electron positron collider, 
operating at centre-of-mass energies of order of 1~TeV, that we shall be able to 
determine the properties of the DM candidate particle and of the other particles 
participating in its interactions in the early Universe, with sufficient accuracy to 
predict the DM relic density precisely. With these results in hand, the comparison 
of the data from CMB experiments, direct DM searches and collider experiments 
would have striking consequences on our quantitative understanding of the 
nature and distribution of dark matter in the Universe.

In these years preceding LHC operation, Supersymmetry has emerged as the best 
motivated theory of new physics beyond the SM. It solves a number of open problems 
intrinsic to the SM and, most important to our discussion, the conservation of 
R-parity introduces the lightest neutralino, $\chi^0_1$, as a new stable, weakly 
interacting particle. CMB data from the WMAP satellite, and other astrophysical data, 
already set rather stringent bounds 
on the parameters of Supersymmetry, if the lightest neutralino is responsible for 
saturating the amount of DM observed in the Universe. The recently released, 
five-year WMAP data provide a determination of the dark matter density as 
$\Omega_{\mathrm{CDM}} h^2$ = 0.110$\pm$0.006~\cite{wmap}.

The potential of the LHC and of an $e^+e^-$ linear collider operating at 
0.5~TeV and 1.0~TeV, such as the International Linear Collider (ILC), 
for determining the neutralino relic density, $\Omega_{\chi}$, in 
Supersymmetry has been investigated in detail in~\cite{Baltz:2006fm}. 
That study selected a set of benchmark points, the so-called LCC points, 
representative of various Supersymmetric scenarios and determined the $\Omega_{\chi}$ 
probability density function by a scan of the full parameter space of the 
Minimal Supersymmetric extension of the SM (MSSM), by retaining those points 
compatible with the measurements available at the LHC and ILC, within their 
experimental accuracy.

In this paper we consider one of the Supersymmetric scenarios defined 
in~\cite{Baltz:2006fm}, for which the neutralino relic density is controlled 
by its annihilation rate through the CP-even heavy Higgs pole $\chi \chi \to A^0$, 
which in turn crucially depends on the value of the mass of the boson, $M_{A^0}$. 
We study the accuracy of the measurement of the relevant properties of the neutral 
heavy Higgs boson $A^0$: its mass, $M_{A^0}$, width, $\Gamma_{A^0}$ and decay 
branching fractions as can be obtained from data collected in  high luminosity 
$e^+e^-$ collisions at centre-of-mass energy of 1~TeV, using full simulation of 
the response of a realistic detector model and detailed event reconstruction.

\section{$e^+e^- \to H^0A^0$ at LCC-4 with Full Simulation}

We adopt the LCC-4 benchmark point of~\cite{Baltz:2006fm}, which is defined in 
the reduced paramater space of the constrained MSSM by $m_0$=380~GeV, 
$m_{1/2}$=420~GeV, $\tan \beta$=53, $A$=0, $Sgn(\mu)$=+1 and $M_{top}$=178~GeV. 
We use {\tt Isasugra 7.69}~\cite{Paige:2003mg} to compute the physical particle 
spectrum and we get $M_{A^0}$=419.4~GeV, $M_{\chi^0_1}$=169.1~GeV and 
$M_{\tilde{\tau_1}}$=195.5~GeV. These parameters correspond to a neutralino 
relic density of $\Omega_{\chi} h^2$ = 0.108, as obtained by using the 
{\tt microMEGAS 2.0} program~\cite{Belanger:2006is}.
The $e^+e^- \to H^0A^0 \to b \bar b b \bar b$ process at $\sqrt{s}$ = 1~TeV
has already been studied for LCC-4 using a parametric
simulation~\cite{Battaglia:2004gk}. We now perform a detailed study using 
{\tt Geant-4}-based full simulation~\cite{Agostinelli:2002hh} of the detector 
response and reconstruct the physics objects using processors developed in 
the {\tt Marlin} framework~\cite{Gaede:2006pj} and extend the analysis to 
both the $b \bar b b \bar b$ and $b \bar b \tau^+ \tau^-$ final states. 
This study adopts the LDC 
detector concept, which is based on a large continuous gaseous tracker, a Time 
Projection Chamber, surrounded  by a highly granular SiW calorimeter and 
complemented by a high resolution Si Vertex Tracker. The LDC detector concept 
is discussed in detail elsewhere\cite{ldc}, the design is optimised for 
achieving excellent parton energy measurements through the 
particle flow algorithm, and precise extrapolation of particle tracks to their 
production point. Both of these features are important to this analysis, which 
aims at suppressing backgrounds by exploiting the signature 4-$b$ and 2-$b$ + 
2-$\tau$ final states of the signal, and requires good determinaton of energy 
and direction of hadronic jets to attain an optimal resolution on di-jet 
invariant mass.

\begin{figure}
\centerline{\includegraphics[width=1\columnwidth]{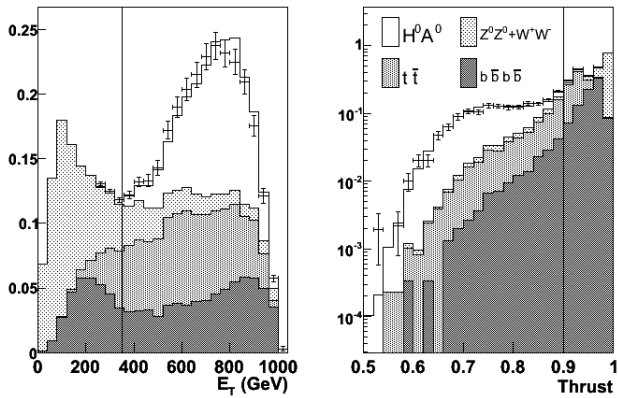}}
\caption{Tranverse energy and thrust distributions for signal and background.
Generator level distributions are plotted as histograms, results of {\tt Mokka +
Marlin} simulation and reconstruction are given for the signal process as points with
error bars. All histograms are normalized to unit area.}
\label{fig:cuts}
\end{figure}

Signal events have been generated with {\tt Pythia 6.205}~\cite{Sjostrand:2000wi} 
+ {\tt Isasugra 7.69}, including beamstrahlung effects~\cite{Ohl:1996fi}. 
At $\sqrt{s}$ = 1~TeV, the effective $e^+e^- \to H^0 A^0$ production cross section, 
accounting for beamstrahlung and initial state radiation, is 1.4~fb, 
BR($A^0 \to b \bar b$) = BR($H^0 \to b \bar b$) = 0.87 and 
BR($A^0 \to \tau^+ \tau^-$) = BR($H^0 \to \tau^+ \tau^-$) = 0.13. 
The main particle pair production backgrounds, $Z^0 Z^0$, $W^+ W^-$ and $t \bar t$, 
have been generated using {\tt Pythia}. Their cross sections, computed using 
{\tt CompHep 4.4.0}~\cite{Boos:2004kh}, are 0.17~pb, 3.0~pb and 0.19~pb respectively. 
The inclusive $b \bar b b \bar b$ and $b \bar b \tau^+ \tau^-$ production, after 
subtracting the contribution of the $Z^0Z^0$ channel and requiring 200~GeV $< M_{bb} <$ 
600~GeV, have cross sections of 0.63~fb and 0.28~fb respectively. These processes have been 
generated at parton level using {\tt CompHep} and then hadronised with {\tt Pythia}.  
We assume to operate the linear collider at $\sqrt{s}$=1~TeV for a total integrated 
luminosity of 2~ab$^{-1}$, which corresponds to 5~years (1~yr = 10$^7$~s) of operation 
for a nominal luminosity of $4 \times 10^{34}$~cm$^{-2}$~s$^{-1}$.

A loose event preselection based on event variables has been applied after 
generation. Selected signal and background events have been passed through the
full LDC simulation using the {\tt Mokka 06-03} program~\cite{Musat:2004sp}, 
an ILC-specific implementation of {\tt Geant-4}. Data are persisted using 
{\tt lcio}~\cite{Gaede:2003ip} collections and used as input for the subsequent
reconstruction in {\tt Marlin}. 

Pattern recognition and track fit are performed first using Monte Carlo truth information 
(``MC truth patrec'') and, for signal events, also genuine full pattern recognition 
(``full patrec''), 
using the {\tt FullLDCTracking} package based on DELPHI experiment software~\cite{Aarnio:1990vx}. 
The performances of these two approaches are compared. The {\tt Pandora v02-00} package 
is used for particle flow~\cite{Thomson:2007zz}. Jet clustering is performed using the DURHAM
algorithm~\cite{Catani:1991hj}. The jet energy resolution has been studied using a 
simulated sample of single $b$ jets in the energy range from 10~GeV to 210~GeV over a 
polar angle, $0.4<\theta<\pi/2 $. We get 
$\delta E/E = {\mathrm{(0.34\pm0.02)}}/\sqrt{E} \oplus {\mathrm{(0.015\pm0.005)}}$,
which is consistent with the LDC particle flow performance specifications.
Jet flavour tagging is performed using the {\tt LCFIVertex} package, which 
developed the original {\tt ZVTOP} tagger~\cite{lcfi} and feeds track and vertex 
topological information into a neural network to distinguish between $b$, $c$ 
and light quark jets. The di-jet mass resolution in the $b \bar b b \bar b$ 
has been improved by performing a constrained kinematic fit.  We have ported 
the {\tt PUFITC} algorithm~\cite{Abreu:1997ic}, developed for the DELPHI 
experiment at LEP2, into a dedicated {\tt Marlin} processor. 
The algorithm adjusts the momenta of the jets given by 
$\vec{p}_{F} =e^{a}\vec{p}_{M}+b\vec{p}_{B}+c\vec{p}_{C}$ where $\vec{p}_{F}$
is the fitted momentum, $\vec{p}_{M}$ is the measured momentum, $\vec{p}_{B}$ 
and $\vec{p}_{C}$ are unit vectors transverse to $\vec{p}_{M}$ and to each 
other, and $a$, $b$ and $c$ are free parameters in the fit.
The adjusted momenta satisfy a set of constraints while minimising the fit 
$\chi^2$, given by 
$\Sigma_i$~$(a_i-a_0)^2/\sigma_a^2 + b_i^2/\sigma_b^2 + c_i^2/\sigma_c^2$, 
where $a_0$ is the expected energy loss parameter, $\sigma_a$ is the energy 
spread parameter and $\sigma_b$, $\sigma_c$ are the transverse momentum spread 
parameters. In this analysis, we impose the following constraints: $p_x=p_y=0$ 
and $E\pm|p_z|=\sqrt{s}$, where the last condition accounts for beamstrahlung 
along the beam axis, $z$.

\subsection{The  $e^+e^- \to H^0A^0 \to b \bar b b \bar b$ Channel}

First we analyse the fully hadronic final state. This provides with 
characteristics four $b$ jet, symmetric events. The backgrounds can be 
significantly suppressed using $b$-tagging, event-shape and kinematic variables. 
We require selected events to fulfill the following criteria: total 
recorded energy in the event $E_{{\mathrm{tot}}}>$ 700~GeV, total transverse energy 
$E_{T}>$350~GeV, total number of reconstructed particles $N_{{\mathrm{tot}}}>$80, number 
of charged particles $N_{{\mathrm{cha}}}>$30, event thrust $<$0.9 and $Y_{34}>$0.002, where 
$Y_{34}$ is the 3 to 4 jet cross-over value of the jet clustering algorithm. The 
distributions of some of these variables are shown in Figure~\ref{fig:cuts} for 
backgrounds and signal, for which a comparison of the generator-level and reconstructed 
values is also given. After event selection, particles are forced into four jets, which are 
arranged into two di-jet pairs, using the pairing which minimises the difference between the 
di-jet masses, $M_{jj}$. The kinematic fit is performed and a cut applied on the resulting 
di-jet mass difference $|M_{jj1}-M_{jj2}|<$50~GeV to eliminate poorly reconstructed events. 
Both di-jet masses are required to satisfy $M_{jj}>$200~GeV.  
The event is required to have four $b$ jets, where a $b$ jet is determined by the following
criteria: total jet multiplicity $N_{{\mathrm{tot}}}>$10, charged jet multiplicity $N_{cha}>$5, 
and b jet probability, $P_b$, larger than 0.5. 
At the chosen working point, an efficiency for $b$ jets of 0.79 is obtained, using 
``MC truth patrec'', with sufficient rejection of lighter quarks to effectively suppress 
the remaining non-$b$ backgrounds. By using ``full patrec'' without retraining  the 
neural net, we measure a tagging efficiency of 0.72 per jet.

\begin{figure}[hb!]
\centerline{\includegraphics[width=1.\columnwidth]{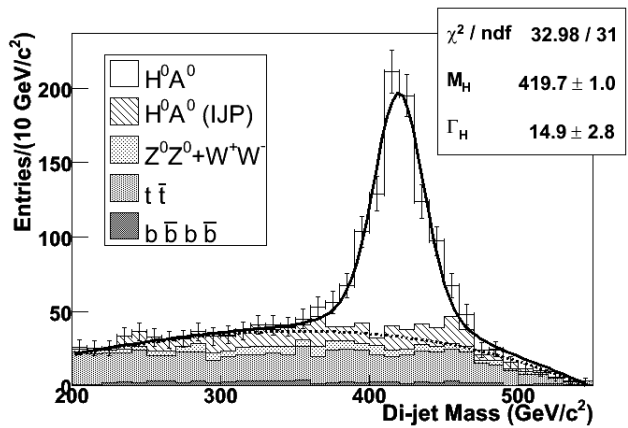}}
\caption{Di-jet invariant mass distribution for signal and background events selected by the 
analysis cuts. Kinematic fit and jet flavour tagging have been applied. $H^0 A^0$ 
events, in which the incorrect jet pairing (IJP) is chosen, are considered as background.}
\label{fig:jjmass}
\end{figure}

The di-jet mass for signal $H^0A^0$ events fulfilling the selection cuts has a Gaussian 
resolution of 23~GeV using tracks reconstructed with ``MC truth patrec'' and 
27~GeV using tracks from ``full patrec'' before the kinematic fit. After applying the 
kinematic fit the di-jet mass resolutions become 13.7~GeV and 13.8~GeV, respectively

After final selection, the sample of events with di-jet 
masses in the region 200~GeV~$< M_{jj} <$~550~GeV gives a selection efficiency for signal 
$b \bar{b} b \bar{b}$ decays of 0.24$\pm$0.01 using tracks reconstructed with ``MC truth patrec'' 
and 0.17$\pm$0.01 using  ``full patrec''. The difference is mostly caused by the observed drop in 
b-tagging efficiency. The corresponding acceptance for 
 $Z^0 Z^0$, $W^+ W^-$, $t \bar t$ and inclusive $b \bar b b \bar b$ background events is
$7 \times 10^{-5}$, $7 \times 10^{-6}$, $8 \times 10^{-4}$ and $4 \times 10^{-3}$, respectively.
The resulting mass distribution is shown in Figure~\ref{fig:jjmass}, which has two entries per event.
The signal is described by the convolution of two Breit-Wigner functions with a mass splitting of 
1.4~GeV, as predicted for the LCC-4 parameters, convoluted with a double Gaussian resolution 
function. The background is described by a third-order polynomial with coefficients determined 
on background only events. 
The final fit function consists of a linear combination of the signal and background functions 
with four free parameters: $M_A$, $\Gamma_A$, and the weights of the signal and background 
functions. We get $M_A$ = (419.7$\pm$1.0)~GeV and $\Gamma_A$ = 
(14.9$\pm$2.9)~GeV, where the quoted uncertainties are statistical only. This result is 
remarkably close to that obtained in the earlier analysis, based on parametric detector simulation.
Using  ``full patrec'' the uncertainties on the $A^0$ boson mass and width increase to 1.3~GeV and
3.4~GeV, respectively.
 
\subsection{The $e^+e^- \to H^0A^0 \to b \bar b \tau^+ \tau^-$ Channel}

\begin{figure}[ht!]
\centerline{\includegraphics[width=1.\columnwidth]{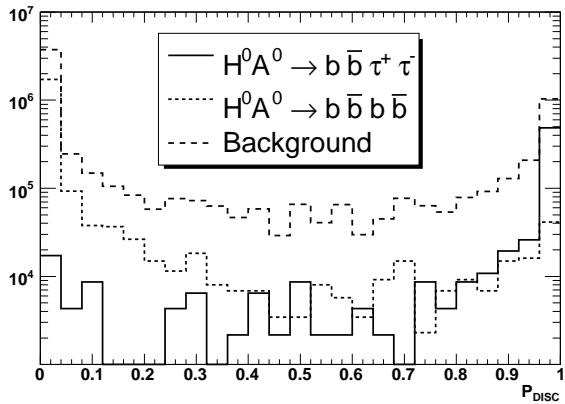}}
\caption{Distribution of the discriminating variable adopted 
for separating $b \bar b \tau^+ \tau^-$ events.}
\label{fig:pdisc}
\end{figure}

The mixed decay mode $b \bar b \tau^+ \tau^-$ can be isolated by tagging a $b \bar b$ di-jet, 
consistent with originating from either a $H^0$ or a $A^0$ decay and analysing the remaining 
particles in the event.
We require the events to fullfill the following criteria: $E_{tot}>$400~GeV, 200~GeV$<E_T<$900~GeV,
40$<N_{tot}<$180, 15$<N_{cha}<$100, event thrust$<$0.8, event sphericity $>$0.1 and $Y_{34}>$0.005.
The event is forced to four jets of which two must be tagged as $b$ jets using the same criteria as 
above but the tighter requirement $P_b >$ 0.9.
The invariant mass of the $bb$ di-jet must satisfy 300~GeV$<M_{bb}<$600~GeV, and that of the two 
remaining jets 250~GeV$<M_{jj}<$600~GeV.  The angle between the two $b$ jets and the angle between 
the two un-tagged jets must satisfy -0.8$<\cos{\theta}<$0. The number of charged particles with 
energy greater than 5~GeV which are not associated to either of the $b$ jets must not exceed six. 
Finally, $\tau$ tagging is performed. We have developed an algorithm 
which outputs a linear discriminant variable $P_{\tau}$ based on the jet mass, the impact parameter 
of the leading track, and a variable, $P_{ISOL}$, which measures the jet energy 
deposited in an annulus around the jet direction. At least one of the two non-$b$ jets must be 
tagged as a $\tau$ jet, where a $\tau$ jet must have less than four energetic charged particles 
and must satisfy $P_{\tau}>$0.8. To distinguish between signal $ b \bar b b \bar b$ and 
$b \bar b \tau^+ \tau^-$ decays, a discriminant variable $P_{DISC}$ is calculated based on the 
un-tagged dijet energy, the number of energetic charged particles not associated to either of the 
two $b$ jets, and $P_{\tau}^{MAX}$, the larger of the two tau jet probabilities (see Figure~\ref{fig:pdisc}). 
The event must satisfy $P_{DISC}>$0.9.
After applying these cuts, the efficiency for signal $b \bar b \tau^+ \tau^-$ decays is 0.14$\pm$0.02,   
that for the $t \bar t$ background is $2\times10^{-4}$, for $Z^0 Z^0$ and $W^+W^-$ is $3\times10^{-8}$
while for $b \bar b b \bar b$ events is $2\times10^{-6}$.
The selection criteria yield 87 events of signal with 89 of background, corresponding to a 
relative statistical uncertainty of 0.15 on the determination of BR($H^0$, $A^0 \to \tau^+ \tau^- $).

\section{Further Constraints on $\Omega_{\chi}$}

The constraints on LCC-4 derived from this determination of the $A^0$ mass and width and 
other supersymmetric particle mass measurements at the LHC and a 1~TeV linear collider, 
provide a prediction of the neutralino relic density with a relative accuracy of 
0.18, within the general MSSM~\cite{Baltz:2006fm}.  
The main contribution to the remaining uncertainty comes from the weak 
constraint which data provide to MSSM solutions where $\Omega_{\chi}$ is 
significantly lower than its reference value for LCC-4. A detailed study shows 
that these solutions are all characterised by large values of the stau trilinear 
coupling, $A_{\tau}$.
In the MSSM the $\tilde \tau$ coupling to the $H^0$ and $A^0$ bosons scales as
$A_{\tau} \frac{\cos \alpha}{\cos \beta} + \mu \frac{\sin \alpha}{\cos \beta}$ 
and $A_{\tau} \tan \beta + \mu$, respectively.
It has been proposed to determine $A_{\tau}$ through a measurement of the 
branching fraction of $A^0$, $H^0 \to \tilde \tau_1 \tilde \tau_2$~\cite{Choi:2005du}.
In the funnel region the main neutralino annihilation mechanism is
$\tilde \chi^0 \tilde \chi^0 \to A^0 \to b \bar b$ and 
$M_A < M_{\tilde \tau_1}+M_{\tilde \tau_2}$. The only $A^0$ decay into 
$\tilde \tau$s allowed by CP symmetry is $A^0 \to \tilde \tau_1 \tilde \tau_2$ 
which is kinematically forbidden for the LCC-4 parameters.
\begin{figure}[ht!]
\centerline{\includegraphics[width=1.\columnwidth]{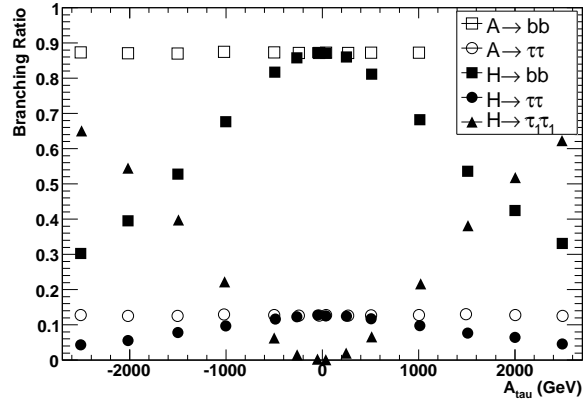}}
\caption{$H^0$ and $A^0$ decay branching fractions as a function of the
stau trilinear coupling $A_{\tau}$ as predicted by {\tt HDECAY}. All the other
MSSM parameters have been kept fixed to those of the LCC-4 point.}
\label{fig:hdecay}
\end{figure}
However, at large values of $|A_{\tau}|$, the 
$H^0 \to \tilde \tau_1 \tilde \tau_1$ decay gets a sizeable enhancement of 
its branching fraction. In this regime, this channel also contributes to the 
neutralino annihilation rate through the 
$\tilde \chi^0 \tilde \chi^0 \to H^0 \to \tilde \tau_1 \tilde \tau_1$ 
process, thus lowering the corresponding relic density, as observed in the 
MSSM scans. At the same time, a determination of the branching fraction
of the decay $H^0 \to \tilde \tau_1 \tilde \tau_1$, allows us to constrain 
the stau trilinear coupling. Figure~\ref{fig:hdecay} shows the decay branching 
fractions of the $A^0$ and $H^0$ bosons computed using the {\tt HDECAY 2.0} 
program~\cite{Djouadi:1997yw} as a function of the $A_{\tau}$ parameter. 
Now, due to the same final state, a large 
$H^0 \to \tilde \tau_1 \tilde \tau_1 \to \tau \tilde \chi^0 \tau \tilde \chi^0$
yield can be detected by a standard $b \bar b \tau \tau$ analysis, such as that 
discussed in the previous section 
The present study shows that the branching fraction for $H^0$, $A^0 \to \tau \tau$ 
can be determined to $\pm$ 0.15 and that for $A^0$, $H^0 \to b \bar b$ to  
$\pm$ 0.07, from which a limit $|A_{\tau}| < $ 250~GeV can be derived. 
This constraint suppresses the tail at low values of $\Omega_{\chi}$ bringing 
the prediction for the neutralino relic density to a relative accuracy of 
0.08, which is comparable to the current accuracy from the WMAP data.

\section{Conclusions}

We have studied the  $e^+e^- \to H^0A^0$ process at $\sqrt{s}$ = 1~TeV using 
on fully simulated and reconstructed events for a Supersymmetric benchmark 
point where the mass of the $A^0$ boson is 419~GeV and the relic Dark Matter 
density in the Universe crucially depends on its mass and width. We find 
that the analysis of 2~ab$^{-1}$ of data should probide with relative 
accuracies of 1.0~GeV and 2.9~GeV in the heavy boson masses and widths, 
respectively. The branching fractions of the $\tau^+\tau^-$ decay can 
be determined with a 0.15 relative accuracy. These data, in combination with 
other measurements available at the LHC and a $e^+e^-$ linear collider, 
allows to infer the neutralino relic density in the Universe with a relative 
accuracy of 0.08. 

\begin{acknowledgments}
We are grateful to Abdel Djouadi for pointing out the sensitivity of the $H$
decay branching fractions to the stau trilinear coupling and to 
Michael Peskin for discussion.
This work was supported by the Director, Office of Science, of the
U.S. Department of Energy under Contract No.DE-AC02-05CH11231 and 
used resources of the National Energy Research Scientific
Computing Center, supported under Contract No.DE-AC03-76SF00098.
\end{acknowledgments}

\end{document}